\documentclass[aps,prl,twocolumn,
               groupedaddress,superscriptaddress,
               amsfonts,amssymb,amsmath,
               citeautoscript,
               a4paper]{revtex4-2}
\usepackage{mathtools}
\usepackage{xcolor}
\usepackage{hyperref}
\hypersetup{colorlinks,
            linkcolor={blue!75!black!80!yellow},
            citecolor={blue!75!black!80!yellow},
            urlcolor={blue!75!black!80!yellow},
            pdfstartview=FitH}
\usepackage{graphicx}
\usepackage{mathrsfs}
\usepackage{amsthm}
\usepackage{physics}
\usepackage{xspace}
\usepackage{braket}
\usepackage{gensymb}
\usepackage{stmaryrd}
\usepackage[UKenglish]{babel}
\usepackage{placeins}
\usepackage{siunitx}
\DeclareSIUnit[number-unit-product=]\percent{\char`\%}
\usepackage{makecell}
\usepackage{diagbox}
\usepackage{float}
\usepackage{empheq}
\usepackage{txfonts}
\usepackage{txfontsb}
\usepackage{accents}

\newcommand*\diff{\mathop{}\mathrm{d}}

\renewcommand{\Re}{\operatorname{Re}}
\renewcommand{\Im}{\operatorname{Im}}
\newcommand{\iu}{\mathrm{i}}

\newcommand{\pt}{\mathcal{PT}}
\newcommand{\U}{\mathcal{U}}
\renewcommand*\Tr{\mathop{}\mathrm{Tr}}

\newcommand{\hkuaffil}{\footnotesize Department of Physics and HK Institute of Quantum Science and Technology, The University of Hong Kong, Pokfulam, Hong Kong, China}
\newcommand{\hkustate}{\footnotesize State Key Laboratory of Optical Quantum Materials, The University of Hong Kong, Pokfulam, Hong Kong, China}

\newcommand*\SMsecCFD{S2}
\newcommand*\SMsecMap{S3}
\newcommand*\SMsecSN{S4}
\newcommand*\SMsecDAC{S6B}

\newcommand*\SMsecSim{S11}
\newcommand*\SMsecPCS{S10}

\newcommand*\SMsecPhase{S9}

\begin{document}

\title{Temporal chirality in Floquet photonic media}

\author{Rui-Chuan Zhang}
\affiliation{\hkuaffil}
\affiliation{\hkustate}
\author{Yixin Sha}
\affiliation{\hkuaffil}
\affiliation{\hkustate}
\author{Yi~Yang}
\email{yiyg@hku.hk}
\affiliation{\hkuaffil}
\affiliation{\hkustate}

\begin{abstract}
We reveal temporal chirality in time-modulated photonic media: reversing the orientation of a closed loop traced by the complex permittivity can leave the Floquet spectrum unchanged while strongly modifying the directional scattering response under the same excitation. This behavior is linked to two symmetry constraints on the evolution operator, namely pseudo-Hermiticity and pseudo-unitarity.
We prove the general occurrence of temporal chirality based on a complex Bogoliubov parametrization of the scattering amplitudes.
We illustrate these findings in a single-harmonic temporal-$\mathcal{PT}$-symmetric system: both modulation-loop orientations support the same vacuum-like Floquet photonic bands at the phase transition, but one suppresses backward photon generation whereas the other induces strong growth, as confirmed by full-wave simulations. Our results establish temporal chirality as a distinct notion in time-varying photonics and identify modulation-loop orientation as a control knob for Floquet wave dynamics.
\end{abstract}

\maketitle

Chirality is a fundamental organizing principle in physics: a system becomes chiral when its response depends on handedness in a way that cannot be removed by any proper rotation. In photonics, chiral behavior is usually associated with spatial structure or propagation~\cite{Peng2016ChiralModesEP}, as in geometrically chiral media~\cite{Rockstuhl2009OpticalActivityChiralMedia,Pendry2004ChiralNegativeRefraction,Gansel2009GoldHelixPolarizer,Tang2010OpticalChirality,Tang2011SuperchiralLight,Yoo2015ChiralResonators,Wang2022MolecularPlasmonics}, structural chirality exhibiting optical spin–momentum locking~\cite{Petersen2014ChiralNanophotonicWaveguide,Lodahl2017ChiralQuantumOptics,VanMechelen2016SpinMomentumLocking}, and topological transport supporting chiral edge modes~\cite{Wang2009TopologicalElectromagneticStates,Skirlo2015LargeChernNumbers,FangYuFan2012NatPhoton,Rechtsman2013PhotonicFloquet,Hafezi2013SiliconTopologicalEdge,Bahari2017NonreciprocalLasing,Wang2021LargeAreaWaveguide}. More generally, the handedness in these phenomena is rooted in the stationary breaking of spatial geometries or time-reversal symmetry, rather than in time ordering itself. This raises a natural question: can an analogous form of handedness arise not from spatial patterning or static bias, but from the temporal sequence in which a material parameter is modulated?

Time-modulated photonic media~\cite{Morgenthaler1958VelocityModulation,Holberg1966ParametricSlab,Galiffi2022PhotonicsTimeVarying,Yin2022FloquetMetamaterials,Asgari2024PTCTutorial,Ptitcyn2023TutorialBasics,Engheta2021HighDegreesFreedom,Engheta2023FourDimensionalOptics} offer a promising platform for this question. Recent studies have shown that such media can induce frequency conversion~\cite{Zhou2020BroadbandFrequencyTranslation,Karl2020FrequencyConversionMetasurface,Moussa2023ObservationTemporalReflection}, parametric amplification~\cite{Lyubarov2022AmplifiedEmissionPTC,Wang2025ExpandingMomentumBandgaps,Xiong2025NonSyntheticPTC,PendryGain2021,Khurgin2024PTCParametricAmplification,GanforninaAndrades2024PRRVacuumAmplification,EchaveSustaeta2024ACSPairCreation,Koufidis2024AlmostPeriodicSlab}, temporal reflection and refraction~\cite{Mendonca2002TimeRefraction,Xiao2014TemporalBoundary,Moussa2023ObservationTemporalReflection,Zhou2020BroadbandFrequencyTranslation,Yin2023ScatteringTemporalInterfaces,Galiffi2025TemporalInterfaces,Zhang2021TemporalReflectionDispersive,Antyufeyeva2025SmoothTimeRefraction}, spatial and temporal PT symmetry~\cite{Li2021TemporalPT,Zhang2025TemporalPTTransition,Feng2017NonHermitianPhotonics,ElGanainy2018NonHermitianPhysics,Chitsazi2017PRL}, and a wide range of related phenomena~\cite{Lustig2018TopologicalPTC,Yin2022TemporalOpticalActivity,Wang2023MetasurfacePTC,Jajin2024ChiralityTimeVarying,Koufidis2024TimePeriodicChiral,Wang2026TemporalFaraday,Lee2026SpontaneousEmissionLasing,Allard2026BroadbandDipoleAbsorption,park2025spontaneousEmissionPTC,Dong2025NonuniformMomentumGap,Galiffi2023PhotonicCollisions,PachecoPena2020TemporalAiming,Li2021TemporalSwitchingAbsorbers,Guo2019NonreciprocalMetasurface,Zang2019TimeModulatedGradient} even in spatially homogeneous systems. 
In particular, when gain and loss modulation are both considered~\cite{Zhang2025TemporalPTTransition,Li2021TemporalPT,Wang2018PhotonicFloquetComplex,PachecoTemporalGainLoss}, the permittivity can trace a closed loop in the complex-permittivity plane rather than a mere line segment along the real permittivity axis, or more generally, an open arc on the complex plane. Reversing the orientation of such a complex loop reverses the temporal ordering with which the field experiences the modulation, thereby providing a natural setting for exploring temporal handedness.

The next question, then, is how temporal chirality appears in measurable response. From the perspective of spectral analysis, periodically driven media are described by Floquet quasienergies, exceptional points~\cite{Doppler2016EncirclingEP,Li2020HamiltonianHopping,Nasari2022ChiralStateTransfer}, and instability windows; such structures can in some cases be reconstructed experimentally~\cite{Park2022NonHermitianFloquetBands,Xiong2025NonSyntheticPTC,Ren2025MomentumGapTopology}. Operationally, however, the system has to be probed through a chosen excitation and the resulting output channels. This distinction suggests a dynamical signature of temporal chirality: two temporal drives---for example, opposite traversals of the same complex-permittivity loop---may share the same Floquet spectrum while producing distinct directional scattering responses under identical excitation.

Here we show that time-modulated photonic media can exhibit a distinct form of temporal chirality. 
This behavior is closely related to the roles played by two symmetry constraints on the evolution operator, namely pseudo-Hermiticity and pseudo-unitarity. 
We prove that temporal chirality can emerge when the evolution operator requires a complex Bogoliubov description beyond its SU(1,1) real-modulation limit.
As an analytically transparent example, we show that even a single-harmonic modulation can produce a pronounced temporally chiral scattering response: at the temporal PT phase transition, a vacuum-like Floquet photonic band emerges for both loop orientations, yet counterclockwise modulation strongly suppresses the generation of backward-scattered photons, whereas clockwise modulation leads to their strong growth.
This asymmetry establishes a directional notion of handedness in photon generation under temporal modulation.

\paragraph*{Modulation loop in the complex permittivity plane.}
We consider a periodic loop \(\Gamma:t\mapsto\epsilon(t)\) in the right half of the complex-\(\epsilon\) plane, so that a continuous branch of \(\sqrt{\epsilon(t)}\) can be well-defined along the cycle.
The traversal orientation of the parametrized closed path \(\Gamma:t\mapsto\epsilon(t)\), with \(\epsilon(0)=\epsilon(T)\), is set by increasing \(t\). Its exact reverse, denoted by \(-\Gamma\), is defined by
\begin{align}
\epsilon_{-\Gamma}(t)
=
\epsilon_{\Gamma}(T-t).
\label{eq:loop_reverse_definition}
\end{align}
We label the two members of a reversal pair by \(\chi_{\Gamma}=\pm1\), with \(\chi_{-\Gamma}=-\chi_{\Gamma}\). We write \(\epsilon(t)=\epsilon'(t)+\iu\epsilon''(t)\), where \(\epsilon'\) and \(\epsilon''\) are the real and imaginary parts of the relative permittivity. For a simple loop, this convention agrees with the signed-area orientation,
\begin{align}
\chi_{\Gamma}
=
\operatorname{sgn}\!\left[
\oint_{\Gamma}
\left(
\epsilon'\diff\epsilon''-\epsilon''\diff\epsilon'
\right)
\right],
\label{eq:loop_orientation}
\end{align}
where \(\chi_{\Gamma}
=+1\) and \(\chi_{\Gamma}
=-1\) correspond to counterclockwise and clockwise traversal, respectively. The path itself need not be simple; self-intersections are allowed because exact reversal is defined by the temporal parametrization rather than by the enclosed area.

Under spatially homogeneous modulation, longitudinal translation symmetry conserves the propagation wave number \(k\equiv k_z\).
Throughout, the permeability \(\mu\) is taken to be time-independent.
For each conserved \(k_z\), we describe the evolution of the normalized field in the forward and backward basis:
\begin{align}
\begin{pmatrix}
\Psi_{+k}^{\mathrm{out}}\\[2pt]
\Psi_{-k}^{\mathrm{out}}
\end{pmatrix}
=
V_k
\begin{pmatrix}
\Psi_{+k}^{\mathrm{in}}\\[2pt]
\Psi_{-k}^{\mathrm{in}}
\end{pmatrix},
\qquad
V_k=
\begin{pmatrix}
f_k & p_k\\
g_k & q_k
\end{pmatrix}.
\label{eq:reduced_U}
\end{align}
For example, for a forward, right-going input \((\Psi_{+k}^{\mathrm{in}},\Psi_{-k}^{\mathrm{in}})^{\mathrm{T}}=(1,0)^{\mathrm{T}}\), \(f_k\) is the forward scattering amplitude while \(g_k\) is the backward left-going scattering amplitude.
For the one-period Floquet operator \(V_k(t_0+T,t_0)\), \(t_0\in[0,T)\) specifies the modulation initial condition, equivalently the starting point along the modulation cycle at which evolution begins.

The Floquet quasienergies are determined by the characteristic polynomial of \(V_k\).
As proved in SM.~\SMsecCFD{} from the Maxwell monodromy, the quasienergy remains invariant under exact reversal of the modulation loop. For the reversed traversal, the modulated permittivity satisfies
\begin{align}
\epsilon_{-\Gamma}(t)=\epsilon_{+\Gamma}(T-t),
\label{eq:loop_reversal_permittivity}
\end{align}
where $\pm\Gamma$ distinguishes the clockwise and counterclockwise orientation of the same modulation loop.
The clockwise and counterclockwise Floquet operators $V_{k}^{-\Gamma}$ and $V_{k}^{+\Gamma}$ both satisfy
\begin{align}
\Tr V_{k}^{+\Gamma} =\Tr V_{k}^{-\Gamma}, \det V_{k}^{\pm\Gamma} = 1,
\label{eq:isospectral_constraints}
\end{align}
Thus $V_{k}^{-\Gamma}$ and $V_{k}^{+\Gamma}$ share the same quasienergy, despite their different matrix elements.

\begin{figure}[t]
    \centering
    \includegraphics[width=1\linewidth]{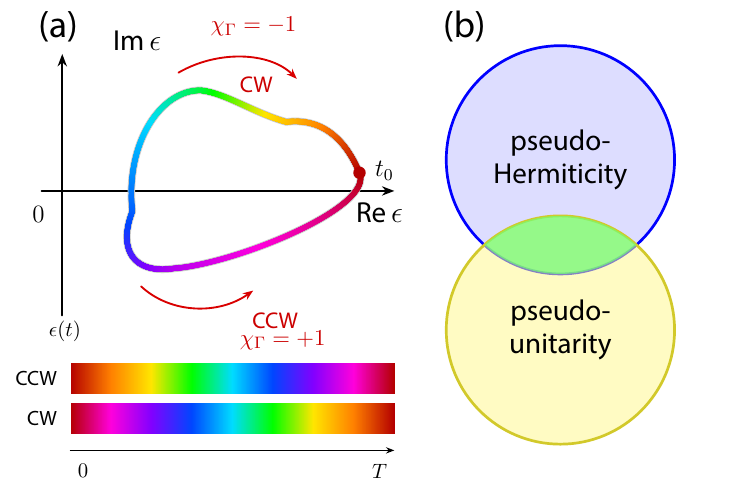}
    \caption{\textbf{Temporal-modulation loop in the complex-permittivity plane and its two symmetry constraints.}
    (a) A closed trajectory \(\Gamma\) in the complex-\(\epsilon\) plane with opposite orientations \(\chi_{\Gamma}=\pm1\). The horizontal colorbars map \(\epsilon(t)\) from \(0\) to \(T\) for the counterclockwise and clockwise traversals, with reversed color order.
    (b) Venn diagram of pseudo-Hermiticity and pseudo-unitarity. The blue circle marks pseudo-Hermiticity, which enforces a real trace of the evolution operator and therefore constrains the Floquet quasienergy. The yellow circle marks pseudo-unitarity, which enforces total momentum conservation of the scattering. The green intersection corresponds to temporal \(\mathcal{PT}\)-symmetric real modulation.}
    \label{fig:PT_ellipse}
\end{figure}

\paragraph*{Pseudo-Hermiticity and pseudo-unitarity.}

Crucially related to temporal chirality, two different symmetries [Figure~\ref{fig:PT_ellipse}(b)] impose different physical constraints on the Floquet operator.

The first symmetry, for a \(\pt\)-invariant choice of the one-period modulation initial condition, is Pseudo-Hermiticity~\cite{Mostafazadeh2002PseudoHermiticity,Zhang2025TemporalPTTransition} (blue circle in Fig.~\ref{fig:PT_ellipse}(b)), under which 
\begin{align}
V_k^{\dagger}=\pt\,V_k\,\pt^{-1}.
\label{eq:spectral_pseudoH}
\end{align}
Here \(\pt\) denotes the temporal-\(\mathcal{PT}\) operator, represented by \(\sigma_y\) in the present forward/backward basis; the general starting-time relation is given in SM.~\SMsecSN. Pseudo-Hermiticity requires the modulation to obey~\cite{Zhang2025TemporalPTTransition}
\begin{align}
    \epsilon_{+\Gamma}(T-t)=\epsilon^{\ast}_{+\Gamma}(t).
    \label{eq:TPT_relation}
\end{align}
Combining Eqs.~\eqref{eq:loop_reversal_permittivity} and~\eqref{eq:TPT_relation} yields
\begin{align}
    \epsilon_{\Gamma}(t)=\epsilon^{\ast}_{-\Gamma}(t)=\epsilon^{\ast}_{\Gamma}(-t).
    \label{eq:TPT_relation_reverse}
\end{align}
Therefore, temporal \(\mathcal{PT}\) symmetry requires the modulation trajectory to be symmetric with respect to the real-\(\epsilon\) axis in the complex-\(\epsilon\) plane.
Taking the trace of Eq.~\eqref{eq:spectral_pseudoH} determines that $\Tr V_k$ must be real.
Pseudo-Hermiticity therefore imposes strong constraints on the quasienergies: they are either purely real or complex conjugate partners~\cite{Mostafazadeh2002PseudoHermiticity}.

Another symmetry at play is the pseudo-unitarity~\cite{Globosits2024PseudounitaryFloquet,OrtegaGomez2023MomentumConservation,Galiffi2022PhotonicsTimeVarying} (yellow circle in Fig.~\ref{fig:PT_ellipse}(b)):
\begin{align}
V_k^{\dagger}\U V_k=\U,
\label{eq:pseudoH_pseudoU}
\end{align}
where $\U=\sigma_z$ in our choice of basis.
We prove in SM.~\SMsecDAC{} that, for the present scalar-$\epsilon(t)$ model with real constant $\mu$, requiring pseudo-unitarity for every $k$ and every modulation initial condition \(t_0\) is equivalent to purely real modulation, i.e., modulation of only $\Re\epsilon$; the complex-$\epsilon$ loop then shrinks into a line segment on the real axis in Fig.~\ref{fig:PT_ellipse}(a).

Contrary to pseudo-Hermiticity, which constrains the quasienergy, pseudo-unitarity imposes longitudinal momentum conservation within the optical fields for this scattering process.
Without loss of generality, again consider the scattering of a forward right-going input \((\Psi_{+k}^{\mathrm{in}},\Psi_{-k}^{\mathrm{in}})^{\mathrm{T}}=(1,0)^{\mathrm{T}}\).
Define the momentum imbalance $\Delta P$
\begin{align}
\Delta P\equiv |f_k|^2-|g_k|^2-1,
\label{eq:fixed_input_momentum}
\end{align}
where the first, second, and third terms correspond to the intensity of the forward scattering wave, backward scattering wave, and the incident wave, respectively.
For temporal modulation obeying pseudo-unitarity, the longitudinal momentum within the optical fields is conserved~\cite{OrtegaGomez2023MomentumConservation} and therefore
\begin{align}
\Delta P=0.
\end{align}
Therefore, regardless of how the waves are amplified or attenuated along the forward and backward directions, the net longitudinal optical-field momentum remains equal to that of the incident wave.
The same reasoning also applies to inputs incident from the backward direction, recalling the usual Bogoliubov constraints \(p_k=g_k^{\ast}\) and \(q_k=f_k^{\ast}\) for real modulation (see SM.~S5).

For generic complex modulation, \(V_k\) generally does not satisfy pseudo-unitarity for all \(k\) and all choices of the modulation initial condition \(t_0\). Complex modulation thus permits
\begin{align}
\Delta P\neq0.
\end{align}
It thus indicates that the longitudinal linear-momentum exchange must happen between the optical fields and the time-modulated medium.
In fact, \(V_k\) can be embedded phenomenologically as the optical-field block of an enlarged canonical field--medium transformation (see SM.~S8).
A generalized momentum conservation can be expressed as
\begin{equation}
\begin{gathered}
\displaystyle V_k\U V_k^{\dagger}+X_k\mathcal U_{\mathrm m}X_k^{\dagger}=\U,\\
\displaystyle V_k^{\dagger}\U V_k+W_k^{\dagger}\mathcal U_{\mathrm m}W_k=\U.
\end{gathered}
\label{eq:completion_balance}
\end{equation}
Here \(\mathcal U_{\mathrm m}=\operatorname{diag}(I_m,-I_m)\) is the canonical signature matrix of a complementary medium subspace containing \(m\) Nambu pairs, while the auxiliary matrices \(X_k\) and \(W_k\) describe the field--medium contribution to the momentum balance~\cite{Caves1982QuantumAmplifiers}.
Equation~\eqref{eq:completion_balance} reduces to the standard pseudo-unitarity Eq.~\eqref{eq:pseudoH_pseudoU} when \(X_k\) and \(W_k\) vanish.
The full temporal chirality response discussed below is already fully encoded in \(V_k\), without the need to specify a microscopic construction of the field--medium exchange.

\paragraph{Complex Bogoliubov analysis of temporal chirality.}
The temporal chirality response can be read from \(V_k\) through a complex extension of the Bogoliubov parametrization (SM.~S6C). 
For nonzero matrix elements, we can express them as
\begin{align}
f_k&=\mathrm e^{\mathrm i\xi_k}\cosh\ell_k, & q_k&=\mathrm e^{-\mathrm i\xi_k}\cosh\ell_k,\nonumber\\
g_k&=\mathrm e^{\mathrm i\zeta_k}\sinh\ell_k, & p_k&=\mathrm e^{-\mathrm i\zeta_k}\sinh\ell_k.
\end{align}
with \(\ell_k,\xi_k,\zeta_k\in\mathbb C\); the determinant-one condition \(f_kq_k-g_kp_k=\cosh^2\ell_k-\sinh^2\ell_k=1\) is automatically satisfied.

Modulation loop reversal provides mappings among these matrix elements. In particular, for integer modulation periods and loop-reversal-invariant initial conditions (SM.~\SMsecMap), exact loop reversal sends \(g_k\mapsto-p_k\) and \(p_k\mapsto-g_k\) (SM.~\SMsecMap). Thus \(|g_k|^2\) and \(|p_k|^2\) are the backward scattering intensities associated with the two opposite loop orientations, respectively. Exact loop reversal also sends \(f_k\mapsto f_k\) and \(q_k\mapsto q_k\), so the forward scattering for an identical input remains unchanged.

It is worth considering the real-$\epsilon$ modulation limit. In this case, the fixed-reference propagator \(V_k(t,t_0)\) lies in $\mathrm{SU}(1,1)$ at arbitrary time $t$~\cite{Yurke1986SU11Interferometers,Truax1985SqueezeOperators} and its parameters \(\ell_k,\xi_k,\zeta_k\) can be chosen as simultaneously real. Therefore, $\ell_k$ is the usual two-mode squeezing parameter, with
\begin{align}
|g_k|&=|p_k|=|\sinh\ell_k|, & |f_k|&=|q_k|=\cosh\ell_k .
\end{align}
As a result, $\sinh^2\ell_k$ and $\cosh^2\ell_k$, respectively, set the identical backward and forward scattering intensities for both modulation loop orientations.

Crucially, complex modulation allows the two backward amplitudes $|g_k|=|\sinh\ell_k|\mathrm e^{-\operatorname{Im}\zeta_k}$ and $|p_k|=|\sinh\ell_k|\mathrm e^{\operatorname{Im}\zeta_k}$ to differ. Although their geometric mean $|\sinh\ell_k|=\sqrt{|g_kp_k|}$ remains unchanged under exact loop reversal, $\operatorname{Im}\zeta_k$ differentiates their relative amplitude and determines which orientation produces stronger backward scattering. In particular, for nonzero \(g_k\) and \(p_k\) we can obtain
\begin{equation}
\operatorname{Im}\zeta_k=\dfrac14\ln\dfrac{|p_k|^2}{|g_k|^2},
\label{eq:chirality_bogoliubov_readout}
\end{equation}
which vanishes for real modulation with \(p_k=g_k^*\). At integer periods, exact loop reversal exchanges \(|g_k|\) and \(|p_k|\), so \(\operatorname{Im}\zeta_k\) flips sign.
This logarithmic intensity ratio in Eq.~\eqref{eq:chirality_bogoliubov_readout} thus provides a direct indicator of temporal chirality manifested in the backward scattering. 

\begin{figure}[htbp]
    \centering
    \includegraphics[width=1\linewidth]{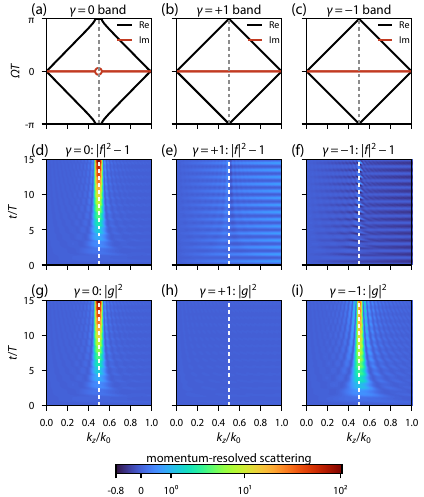}
    \caption{\textbf{Isospectral bands and their distinct momentum-resolved directional scattering.}
    {
    (a--c) Floquet quasienergy bands when \(\gamma=0,+1,-1\) for Eq.~\eqref{eq:phase_lag_family}, respectively, with real and imaginary parts shown by black and red curves. Panels b and c are isospectral. The vertical dashed line marks the primary parametric resonance at \(k_z/k_0=0.5\). 
    (d--f) Momentum-resolved forward scattering \(|f|^2-1\), and (g--i) backward-scattering \(|g|^2\), over \(0\leq t/T\leq15\) and \(0\leq k_z/k_0\leq1\).
    Identical forward (d) and backward (g) scattering appears for the real modulation limit.
    When \(\gamma=\pm1\), \(|f|^2-1\) coincides at integer periods in panels (e,f).
    Here \(\epsilon_{\mathrm{i}}=2\) and \(\epsilon_{\mathrm{m}}=0.5\).  
    }
    }
    \label{fig:mode_resolved_conversion}
\end{figure}

\paragraph*{Temporal chirality under single-harmonic modulation.}
To see temporal chirality explicitly, we now focus on a single-harmonic temporal \(\mathcal{PT}\)-symmetric modulation as an analytically tractable example:
\begin{align}
\epsilon(t;\phi,\gamma)
=
\epsilon_{\mathrm{i}}
+
\epsilon_{\mathrm{m}}
\left[\cos(\omega_{\mathrm{m}}t)+\iu\gamma\cos(\omega_{\mathrm{m}}t-\phi)\right],
\label{eq:phase_lag_family}
\end{align}
where \(\epsilon_{\mathrm{i}}>0\), \(\epsilon_{\mathrm{m}}\), and \(\omega_{\mathrm{m}}\) are real. Here \(\phi\) is the phase delay, and \(\gamma\) is the contrast between the imaginary and real modulation depth.
\(\gamma=0\) is the real-modulation limit, and for \(0<\phi<\pi\) the sign of \(\gamma\) directly corresponds to the modulation loop orientation (Fig.~\ref{fig:PT_ellipse}(a)), i.e., \(\chi_{\Gamma}=\operatorname{sgn}\gamma\).
The temporal-\(\mathcal{PT}\) phase transition occurs at \(\phi=\pi/2\) and \(\gamma=\pm1\).

Fig.~\ref{fig:mode_resolved_conversion} shows the Floquet bands and the analytically calculated forward and backward scattering under modulation with $\gamma=0$, $\gamma=1$, and $\gamma=-1$, respectively.
In the real modulation limit of $\gamma=0$, a momentum gap is opened centered at $k_z=0.5k_0$, where \(k_0\equiv\sqrt{\epsilon_{\mathrm i}\mu}\,\omega_{\mathrm m}/\mathrm c\) (Fig.~\ref{fig:mode_resolved_conversion}(a)), leading to exponential amplification of both forward and backward scattering in Fig.~\ref{fig:mode_resolved_conversion}(d) and (g).
For $\gamma=\pm1$, their associated Floquet bands are identical (Fig.~\ref{fig:mode_resolved_conversion}(b) and (c)), and their forward responses coincide after every integer modulation period, although their intra-period evolution could differ (Fig.~\ref{fig:mode_resolved_conversion}(e) and (f)). 
The oscillations between successive integer periods arise from Floquet micromotion~\cite{Goldman2014Micromotion,Bukov2015FloquetReview}.

Importantly, their backward scattering differs substantially: almost negligible backward scattering appears for $\gamma=1$ (Fig.~\ref{fig:mode_resolved_conversion}(h)) whereas strong backward scattering amplification is exhibited for $\gamma=-1$ (Fig.~\ref{fig:mode_resolved_conversion}(i)).
Modulation loop reversal interchanges the \(\pm\omega_{\mathrm{m}}\) Fourier components of the modulation. Near the primary parametric resonance, the \(\gamma=+1\) orientation lacks the leading-order resonant harmonic in the backward-scattering channel, whereas the \(\gamma=-1\) orientation retains it and produces strong backward scattering (see SM.~\SMsecPCS{}).
More quantitatively, for weak modulation, the one-period
backward scattering intensity under the temporal $\mathcal{PT}$
condition $\phi=\pi/2$ admits, at the primary parametric resonance
$2\omega_{k,t_0}=\omega_{\mathrm{m}}$ (where \(\omega_{k,t_0}\) is the mode frequency at the modulation initial condition \(t_0\)) the asymptotic form
\begin{align}
|g_k|^2
=
\dfrac{\pi^2}{16}
m_\epsilon^2(\gamma-1)^2
+O(m_\epsilon^3),
\label{eq:quadratic_backward_minimum}
\end{align}
where $m_\epsilon\equiv\epsilon_{\mathrm{m}}/\epsilon_{\mathrm{i}}\ll1$
is the dimensionless modulation depth. Thus, the leading-order backward scattering intensity has a quadratic
minimum for the counterclockwise modulation $\gamma=1$,
whereas the leading resonant contribution remains finite for
the clockwise modulation $\gamma=-1$
(see SM.~\SMsecPCS{}).

\begin{figure}[htbp]
    \centering
    \includegraphics[width=1\linewidth]{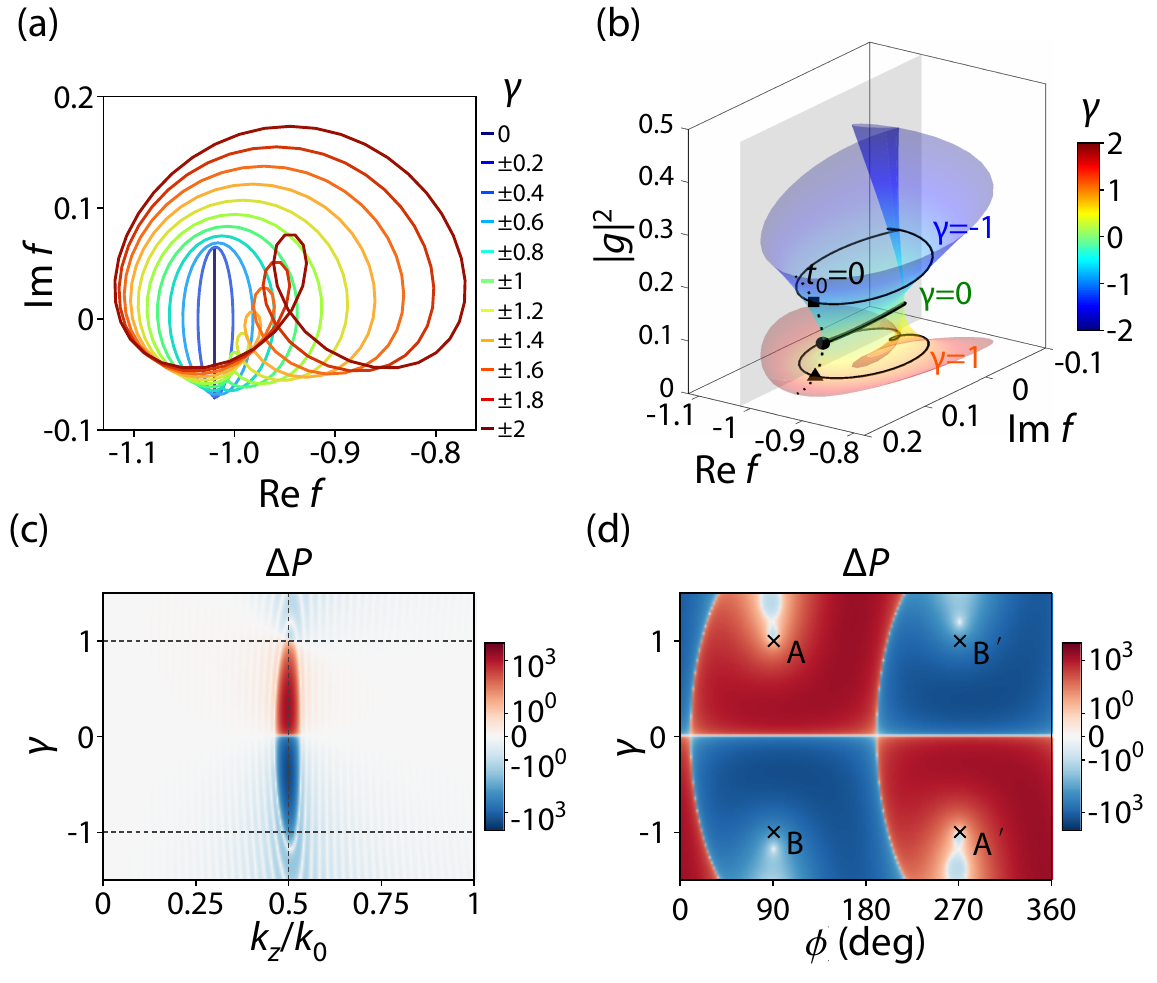}
    \caption{\textbf{Matrix element evolution and momentum imbalance as a function of modulation depth contrast $\gamma$ and phase delay $\phi$.}
    (a--b) Evolution operator's matrix-element trajectories in the \((\Re f,\Im f)\) plane (a) and one-period matrix-element trajectories in the \((\Re f,\Im f,|g|^2)\) manifold (b) as a function of $\gamma$ for the single-harmonic modulation in Eq.~\eqref{eq:phase_lag_family} at \(k_z/k_0=0.5\), with \(t_0\) varied over one period.
    In (a), each \(\pm\gamma\) pair traces the same projected loop in opposite directions.
    In (b), the solid curves label \(\gamma=0,\pm1\), and the dashed curve marks \(t_0=0\).
    (c) Momentum imbalance \(\Delta P\) for a forward-going input $(1,0)^\mathrm{T}$ as a function of \(k_z/k_0\) and \(\gamma\), evaluated for \(t_0=0\) after \(N=25\) periods.
    (d) Momentum imbalance \(\Delta P\) as a function of \(\phi\) and \(\gamma\) at \(k_z/k_0=0.5\), \(t_0=0\), and \(N=25\).
    Numerical panels use \(\epsilon_{\mathrm{i}}=2\), \(\epsilon_{\mathrm{m}}=0.5\).
    }
    \label{fig:momentum_imbalance}
\end{figure}

To further understand this distinction, we consider the trajectory of the matrix elements in Fig.~\ref{fig:momentum_imbalance}.
Fig.~\ref{fig:momentum_imbalance}(a) illustrates the evolution of $\Re{f}$ and $\Im{f}$ for various $\gamma$. For a $\pm\gamma$ pair, the $\Re{f}$-$\Im{f}$ trajectories are degenerate yet show opposite looping orientations. 
Also notably, at $\gamma=0$, the loops collapse into a line segment.
Furthermore, temporal-\(\mathcal{PT}\) symmetry imposes the constraint that \(f_{k,\gamma}(T-t_0)=\Tr V_{k,\gamma}-f_{k,\gamma}(t_0)^{\ast}\). Hence each projected $\Re{f}$-$\Im{f}$ trajectory is mirror-symmetric about the \(\Re f=\Tr V_{k,\gamma}/2\) axis that relates to the Floquet spectra.
The two \(\mathcal{PT}\)-invariant modulation initial conditions, \(t_0=0\) and \(t_0=T/2\) modulo \(T\), lie on this symmetry axis, satisfying \(\Re f_{k,\gamma}=\Tr V_{k,\gamma}/2\).
The joint dependence of the backward scattering $|g|^2$ as a function of $\gamma$ and $f$ is shown in Fig.~\ref{fig:momentum_imbalance}(b).
Consistent with Fig.~\ref{fig:mode_resolved_conversion}(h) and (i), a negative looping orientation ($\gamma<0$) generally produces much stronger backward scattering than that of a positive looping orientation $\gamma>0$, highlighting the occurrence of temporal chirality.

Because pseudo-unitarity is broken under complex modulation, next we compare the momentum imbalance $\Delta P$ (see Eq.~\eqref{eq:fixed_input_momentum}) under different chirality. 
Fig.~\ref{fig:momentum_imbalance}(c) shows a pronounced scattering response within the momentum-gap region, where forward and backward scattering dominate for either $0<\gamma<1$ or $-1<\gamma<0$, respectively; see the red and blue tip-like regions. Separating the two regions at $\gamma=0$ is a clear boundary of momentum conservation, where pseudo-unitarity is obeyed. 
Furthermore, it is worth emphasizing that the momentum imbalance is not mirror-antisymmetric with regard to $\gamma=0$: for both $\gamma>1$ and $\gamma<-1$, the backward scattering outweighs the forward one.

So far, we have fixed the relative phase between the reactive and dissipative modulations at \(\phi=\pi/2\). We now release this constraint and vary \(\phi\) as a continuous control parameter while keeping the mode at the center of the momentum gap, \(k_z/k_0=0.5\). 
Figure~\ref{fig:momentum_imbalance}(d) maps the resulting momentum imbalance $\Delta P$ over the \((\phi,\gamma)\) parameter space. 
Because the signed area of the temporal loop is exactly \(\dfrac{1}{2}\oint_{\Gamma}(\epsilon'\diff\epsilon''-\epsilon''\diff\epsilon')=\pi\epsilon_{\mathrm{m}}^2\gamma\sin\phi\) for the modulation scheme in Eq.~\eqref{eq:phase_lag_family},
either shifting \(\phi\) by \(\pi\) or reversing the sign of \(\gamma\) interchanges the $(\phi,\gamma)$ patches dominated by either forward scattering (\(\Delta P>0\)) or backward scattering (\(\Delta P<0\)). 
In between these patches, the horizontal boundary is the real modulation line \(\gamma=0\) satisfying longitudinal momentum conservation within the optical fields \(\Delta P=0\).
The additional near-vertical curved boundary is a parameter-dependent and input-specific balance contour (SM.~\SMsecPhase{}) rather than a symmetry-imposed boundary. 

The four temporal-\(\mathcal{PT}\) points \((\phi,\gamma)=(\pm\pi/2,\pm1)\), labeled as A/A' and B/B' in Fig.~\ref{fig:momentum_imbalance}(d), are worth more elaboration.
They share the same Floquet spectra under loop reversal, but exhibit contrasting scattering dynamics (see numerical details in SM.~\SMsecPhase) and substantially different momentum imbalance $\Delta P$.
Mathematically,
the A and A' suppression of backward scattering follows from the absence of the leading-order resonant pump harmonic in the backward channel, whereas such leading-order contribution remains finite in the backward channels at the B and B' points [see Eq.~\eqref{eq:quadratic_backward_minimum}].

\begin{figure}[htbp]
    \centering
    \includegraphics[width=1\linewidth]{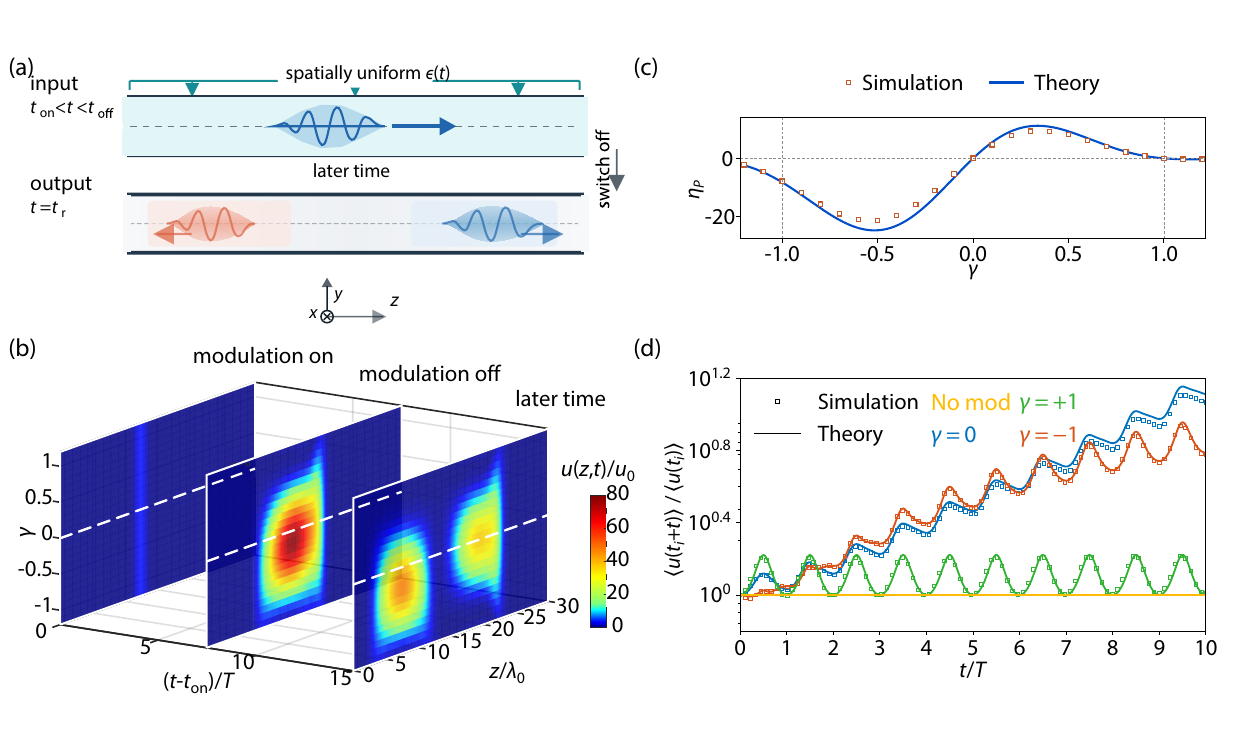}
    \caption{\textbf{Full-wave numerical verification of temporal chirality.}
    (a) A $y$-polarized Gaussian pulse enters a homogeneous time-modulated medium from the left. After the modulation ends, the forward and backward waves are integrated over the right and left halves of the simulation domain, respectively.
    (b) Normalized energy-density distributions \(u(z,t)/u_0\) at modulation switch-on, switch-off, and a later time for different $\gamma$. Here \(u_0=\epsilon_0E_0^2/2\) is the energy-density normalization.
    (c) Spatially integrated momentum imbalance $\eta_P$ as a function of $\gamma$ for a 15-period modulation, $t_{\rm off}-t_{\rm on}=15T$.
    (d) Normalized spatially integrated energy-density evolution for $\gamma=\left\{0,\pm1\right\}$ and the unmodulated case. In (c) and (d), open squares and solid lines denote the simulations and Floquet theory, respectively.
    }
    \label{fig:comsol}
\end{figure}

\paragraph*{Full-wave simulations.} 
We verify temporal chirality using full-wave time-domain simulations. Figure~\ref{fig:comsol}(a) provides a schematic of the simulation setup. A $y$-polarized Gaussian pulse is launched from the left into a spatially homogeneous medium whose permittivity is uniformly modulated for $t_{\mathrm{on}}<t<t_{\mathrm{off}}$. After the modulation is switched off, the forward and backward waves propagate in opposite directions and become spatially separated.

For the full-wave analysis, we consider the evolution of time-domain instantaneous energy density (see SM.~S11B):
$
u(z,t)=\dfrac{1}{2}\epsilon_0\Re{\epsilon(t)}|E_y(z,t)|^2+\dfrac{1}{2}\mu_0\mu|H_x(z,t)|^2,
$
where we take \(\mu=1\) and assume a local nondispersive treatment of the media.
Fig.~\ref{fig:comsol}(b) shows three snapshots of the normalized energy-density distribution \(u(z,t)/u_0\), taken when the modulation is switched on, switched off, and at a later time. At the later time, the backward wave is enhanced for negative $\gamma$ and suppressed for positive $\gamma$, revealing a clear dependence on the temporal-loop orientation. 

To quantify this behavior, we integrate $u(z,t)$ over the spatially separated forward and backward scattered waves and average over a short post-modulation window \(\mathcal T_{\mathrm{eval}}=[t_1,t_2]\). We divide the spatial domain into the right half \(\mathcal D_+\) and the left half \(\mathcal D_-\), and define the time-averaged spatially integrated energy per unit transverse area as
$
    P_\pm=\frac{1}{t_2-t_1}\int_{t_1}^{t_2}\!\mathrm{d}t\int_{\mathcal D_\pm}u(z,t)\,\mathrm{d}z
$.
This further allows us to define a spatially integrated momentum imbalance \(\eta_P\equiv(P_+-P_--P_{\mathrm{inc}})/P_{\mathrm{inc}}\) (SM.~\SMsecSim{}), where \(P_{\mathrm{inc}}\) is the spatially integrated incident-pulse energy per unit transverse area before the modulation starts. Here \(P_+\), \(P_-\), and \(P_{\mathrm{inc}}\) are energy integrals, so \(\eta_P\) is an energy-based proxy for the normalized optical momentum imbalance \(\Delta P\).
The particular choice of $t_1$ and $t_2$ does not matter as long as the temporal modulation has ended and the two counter-propagating scattered waves are spatially separated (see SM.~\SMsecSim).

\(\eta_P\) can be understood as a spatially integrated version of momentum imbalance \(\Delta P\).
Fig.~\ref{fig:comsol}(c) plots $\eta_P$ as a function of $\gamma$ and confirms the pronounced difference between the two loop orientations (as also predicted by Floquet theory). Its pulse-integrated normalization is detailed in SM.~\SMsecSim.
The field evolution further reveals distinct amplification dynamics. In Fig.~\ref{fig:comsol}(d), the Gaussian-pulse simulations and Floquet theory consistently show net amplification for $\gamma=0$ and $\gamma=-1$, but no sustained growth for $\gamma=+1$.

\paragraph{Conclusions.}
\label{sec:conclusion}

We identified and characterized a distinct form of temporal chirality in Floquet photonic media. The handedness of a modulation protocol is not exhausted by its Floquet spectrum: opposite temporal traversals of the same complex-permittivity loop can remain Floquet-isospectral while producing sharply different scattering responses under the same excitation. In this sense, temporal chirality is revealed through directional scattering dynamics, rather than solely through spectral diagnostics. This establishes a clear distinction between spectral equivalence and dynamical equivalence in periodically driven photonic systems. 

More broadly, our results suggest that time modulation can encode a directional degree of freedom that is neither reducible to ordinary spatial chirality nor to static nonreciprocal bias. Instead, the relevant asymmetry is carried by the ordering of the modulation history itself. 
This viewpoint expands the conceptual scope of chirality in photonics and points toward a broader class of temporal control strategies in which loop orientation emerges as a new complementary degree of freedom---alongside modulation amplitude, frequency, and phase---for wave manipulation.

\paragraph{Acknowledgement.}
We thank Joseph Huang, Mordechai Segev, Qingdong Yang, and Shuang Zhang for helpful discussions.

\input{main23.bbl}

\end{document}